\documentstyle[psfig,11pt]{article}
\begin{document}
\vspace{1.0cm}
{\Large \bf Detection of Iron Emission Line from the Galaxy Cluster 
Including the Distant Radio Galaxy 3C220.1}

\vspace{1.0cm}

Naomi Ota$^1$, Kazuhisa Mitsuda$^1$, Makoto Hattori$^2$ and Tatehiro 
Mihara$^3$

\vspace{1.0cm}
$^1${\it Institute of Space and Astronautical Science,
         3-1-1 Yoshinodai, Sagamihara, Kanagawa, 229-8510, Japan }\\
    \centerline{ota@astro.isas.ac.jp}\\
$^2${\it Astronomical Institute, T\^{o}hoku University, Aoba, Sendai, 980-8578, Japan}\\
$^3${\it Institute of Physical and Chemical Research, 2-1 Hirosawa, Wako, 
Saitama, 351-0198, Japan}\\

\vspace{0.5cm}

\section*{ABSTRACT}
We detected an emission line feature at 4 keV in the X-ray spectrum of
a sky region including the distant radio galaxy 3C220.1($z=0.62$)
obtained with ASCA.  The line energy is 6.1 -- 7.0 keV (90\%
confidence) in the rest frame of 3C220.1.  Within the present
statistics, the observed spectra are consistent with two different
models; a non-thermal model consisting of a power-law continuum plus a
6.4 keV iron emission line, and a Raymond-Smith thin-thermal emission
model of $kT \sim 6$ keV with a metal abundance of $\sim0.5$
solar. However, the large ($\sim 500$ eV) equivalent width of the line
indicates that a significant fraction of the X-ray emission is likely
to arise from the hot intracluster gas associating the galaxy cluster
including 3C220.1. The spectral parameters of the thermal emission are
consistent with the luminosity-temperature relation of nearby clusters
and the mass estimates from the giant luminous arc.

\section{INTRODUCTION}
Among a variety of attempts searching for clusters at high redshifts,
X-ray observation directly illuminates the gravitational potential
well because the X-ray emitting hot plasma is considered to trace
it. Moreover X-ray emission lines from highly ionized ions, irons in
particular, can be used to determine the redshift of the hot gas,
thereby its association with other objects can be investigated.  Based
on these ideas we have conducted an observation of the radio galaxy
3C220.1 with ASCA.

An existence of a galaxy cluster which surrounds the radio galaxy
3C220.1 ($z=0.62$) was first suggested from the observations at Lick
Observatory (Dickinson, 1984). Around the radio galaxy in the Lick
image, the presence of several quite red galaxies were found. Further
observations were performed at Kitt Peak National Observatory, and the
blue band image revealed the presence of a giant luminous arc
(Dickinson, 1984).  In 1995, much higher quality images were obtained
with Hubble Space Telescope (Dickinson, 1998). A large arc (9 arc
seconds in radius and subtending $\sim$ 70 degrees around the radio
galaxy) was clearly resolved.

The arc image of 3C220.1 is very symmetric and regarded as a section
of an Einstein ring caused by gravitational lensing. This offers a
remarkable tool to constrain the cluster mass enclosed within it.  The
redshift of the arc was successfully determined using the Keck
telescope to be $z_s=1.49$ (Dickinson, 1998). Under an assumption of
spherically symmetric geometry, the projected lensing mass within the
arc radius, $9"$, is $M_{\rm lens}(<9") = 3.8\times10^{13}{\rm
M_{\odot}}$. Here $\Omega_0=1$, $\Lambda=0$, and $H_0=50 {\rm \,km\,
sec^{-1}Mpc^{-1}}$ are adopted. The derived mass is appropriate to
clusters of galaxies rather than a single galaxy.

In this paper we report the detection of an iron-K emission line with 
ASCA and discuss the origin of the X-ray emission.

\section{OBSERVATION AND RESULTS}

\subsection{\underline{Observation}}
We observed 3C220.1 with the ASCA GIS and SIS for 40 ksec on 1998
April 11, in AO6 period. The GIS was operated in the PH nominal mode,
while the SIS was in the Faint 1CCD mode. The data were filtered by
the standard ASCA data screening procedure.

The X-ray emission centered at $(9^{\rm h} 32^{\rm m} 43^{\rm s},
+79^{\circ} 06' 39") _ {\rm J2000.0}$ is detected.  The peak position
is about 0.2 arcmin off from the cataloged value of 3C220.1, however
is consistent with it within the uncertainty of ASCA attitude
determinations (0.5 arcmin at 90\% confidence). In the GIS and SIS
fields, there are several other bright sources in the vicinity of
3C220.1.  Most of them are not identified with known objects, except
for one at $(9^{\rm h} 31^{\rm m} 34^{\rm s}, +79^{\circ} 04' 9")_{\rm
J2000.0}$ which is a radio-loud AGN (see comments in Hardcastle et al.
1998).

\subsection{\underline{Spectral Analysis}}

Since the angular separations between 3C220.1 and three nearby
sources are 3.4 to 5.1 arcmin, contamination of 3C220.1 energy
spectrum from these sources must be carefully treated. In order to
check the contribution due to the nearby sources, we have accumulated
the energy spectra in two different integration regions: (1) a
circular region centered on 3C220.1 with a radius of $3.0'$ and (2)
the same circular region as (1) but excluding three circular areas centered
on the nearby sources. The radii of the excluded regions are
proportional to the intensity of the nearby sources. We then evaluated
these two spectra by model-fittings. For both  power-law model or
 Raymond-Smith model, the resultant model parameters are consistent
with each other within the statistical errors. In what follows, we
show the results for case (1) because at present only the azimuthally
averaged response function is available for the X-ray telescope, thus
case (2) may involve some systematic errors.

Firstly we tried a single power-law model with a neutral absorption
with a column density fixed at the Galactic value,
$N_{\rm H}=1.93\times10^{20} {\rm cm^{-2}}$ (Stark et al. 1992).
 The absorption is fixed
to this value throughout this paper. The fits are acceptable with a
best-fit power-law photon index of $1.9 (^{+0.1}_{-0.2}$, 90\% error),
however, the fit leaves two excess data points at around 4 keV
for the SIS and GIS spectra. If we add a narrow Gaussian
emission line, the fit improves from the $\chi^2$ value of 33.9 for 34
degrees of freedom to 26.4 for 31 degrees of freedom. The improvement
is significant by the F-test at a 95\% confidence limit. 

The best-fit Gaussian central energy is 3.9 keV in our frame with the
90\% error range of $3.8-4.2$ keV. The most likely origin of this line
feature is a red-shifted iron emission line.  If we assume
low-ionization iron emission lines at 6.4 keV, the redshift is
estimated to be 0.63($0.53-0.69$, 90\% error), while if we assume 6.7
keV lines from Helium-like irons, the redshift is $0.71 (0.61-0.76)$.
Therefore the redshift of the radio galaxy ($z=0.62$) is within the
error range in either case.

If we assume that the line emission is originating from an object at
$z=0.62$, the central energy is estimated to be $6.3 (6.1-7.0)$ keV in
the rest frame. It can be interpreted either as 6.4 keV low-ionization
iron emission line which may be associated with the AGN, 
6.7/6.9 keV lines
from highly ionized irons which may be attributed to cluster hot gas,
or a combination of these two. We are not able to distinguish these
emission lines under the current limited statistics and the detector
resolutions.
 
\begin{figure}[htb]
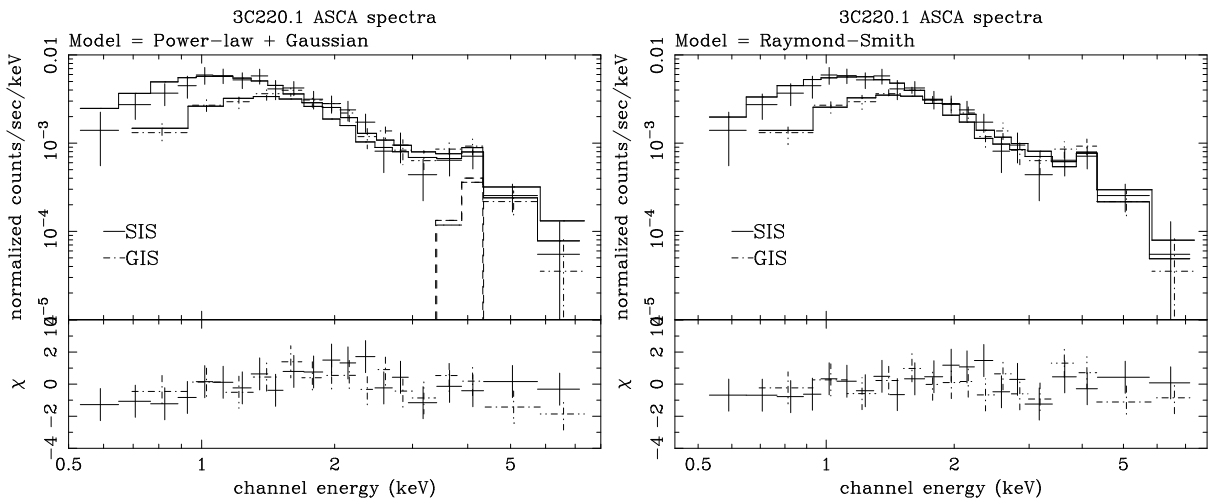

\centerline{
\psfig{figure=const.ps,height=6.5cm,angle=270}
\psfig{figure=raymond2.ps,height=6.5cm,angle=270}
}
\caption[]{ASCA SIS(0+1) and GIS(2+3) spectra fitted with the power-law plus 
Gaussian model(left panel), and the Raymond-Smith model(right panel). 
The crosses denote the observed spectra and the step functions show
the best-fit model function convolved with the X-ray telescope and the
detector response functions.}
\label{fig:spec}
\end{figure}

Thus next, we performed fits with models corresponding to the above
two cases; power-law model plus a Gaussian with line center energy
fixed at 6.4 keV and Raymond-Smith model representing an optically
thin thermal plasma emission. In Figure \ref{fig:spec} and Table
\ref{tab:result}, the results of the fits are shown, where the model
parameters for the GIS and SIS spectrum are combined except for their
normalization factors. Although both models are acceptable at the 90\%
confidence limit, the Raymond-Smith model gives a  smaller
reduced $\chi^2$ value.

\begin{table}[htb]
\begin{center}
\caption[]{Results of Spectral Fits}

\label{tab:result}
\begin{tabular}{lll}\hline\hline
Model & Parameter & Value(error$^{a}$) \\\hline 
 Power-law plus Gaussian &Photon Index & $1.9 (1.8-2.1)$\\ 
			&Equivalent width[eV]$^{b}$& $480(190-780)$\\
			&$L_{\rm X}^{2-10}$[erg/s]$^{c}$&$8.4\times10^{44}$\\
			&$\chi^2$/d.o.f&26.7/32\\\hline 
 Raymond-Smith &$kT$[keV] & $5.6(4.5-7.1)$\\ 
		&Abundance[Z$_{\odot}$] &$0.54(0.17-1.0)$\\
			&$L_{\rm X}^{2-10}$[erg/s]$^{c}$&$8.4\times10^{44}$\\
		&$\chi^2$/d.o.f&18.0/33\\\hline 
\end{tabular}
\end{center}
The absorption column density is fixed at the Galactic value;
$N_{\rm H}=1.93\times10^{20}{\rm cm^{-2}}$.  $^{a}$The quoted errors
correspond to a single parameter error at 90\% confidence.
$^{b}$A narrow line at 6.4 keV is assumed.
$^{c}$Absorption corrected $2-10$ keV luminosity assuming the distance of $z=0.62$.
\end{table}

\section{DISCUSSION}
We have detected an emission line at 4 keV (3.8 -- 4.2 keV) in our
rest frame, which corresponds to 6.1 -- 7.0 keV at $z=0.62$.  Within
the present statistics, the observed spectra are consistent with two
different models; a non-thermal model which consists of a power-law
continuum and a 6.4 keV emission line, and a Raymond-Smith thermal
model with temperature about 6 keV. In this section, we discuss the
origin of the X-ray emission.

The radio source 3C220.1 is classified as an FRII narrow emission line
galaxy (NELG).  Turner et al.(1997) investigated the narrow iron K
line at 6.4 keV from type 2 AGNs systematically, and reported most of
the NELG shows an equivalent width smaller than 200 eV and on average
$\sim$ 100 eV. Thus for 3C220.1, the derived equivalent width under a
non-thermal model is a factor of 2 to 8 larger than the typical
NELGs if one attributes all of the line intensity to the radio galaxy
3C220.1.  This indicates a large ($ > 80 $\%) fraction of the iron
line is emitted from the other emission region; most likely
intracluster medium of the galaxy cluster.

The ROSAT HRI observations revealed that the X-ray emission consists
of an extended ($\sim 10"$) component which carries about 60\% of the
HRI photons and a compact central component (Hardcastle et al.  1998).
Since two spectral models in our analysis are not distinguished within
the statics, any combinations of the two models should also be
statistically acceptable.  If the compact component carries 40\% of
photons in the ASCA energy band and contains an iron emission line of
100 eV equivalent width in its own spectrum, the equivalent width for
the rest of the emission is expected to be 150 eV to 1100 eV. This
value is appropriate for a thin thermal emission of $kT\sim6$ keV with
metal abundance of 0.2 to 1.3 solar .

If we consider about 60\% of the total emission is arising from the
cluster, the luminosity and the temperature obtained from the spectral
fit are consistent with the luminosity-temperature relation for nearby
clusters obtained by David et al.(1993) within the scatter of data
points.

The masses of various clusters determined from the gravitational arc (lens
mass) and determined from the X-ray observations (X-ray mass) have
been compared by several authors (e.g. Wu\&Fang 1997). These results
shows that in general the X-ray mass becomes either consistent with the
lens mass or smaller.  Adopting the $\beta$ model parameters obtained
from the ROSAT HRI observations ($\beta=0.9$ and the core
radius, $r_c=13"$), we obtain that the X-ray mass becomes equal or
smaller than the lens mass if the temperature of the X-ray emission is
lower than 6.4 keV. The determined X-ray temperature of 
$5.6^{+1.5}_{-1.1}$ keV from the single Raymond-Smith model fits 
suggests this is the case.

In conclusion, a large fraction of the iron line emission and the
continuum emission associated with the line is likely to originate
from the intracluster medium in the galaxy cluster around
3C220.1. This is a strong evidence for the existence of the cluster of
galaxies including 3C220.1.

\section{REFERENCES}
\vspace{-5mm}
\begin{itemize}
\setlength{\itemindent}{-8mm}
\setlength{\itemsep}{-1mm}
\item[]
David, L.P., Slyz, A., Jones, C., Forman, W., and Vrtilek, S.D., 
{\it ApJ}, {\bf 412}, 479 (1993).
\item[] 
Dickinson, M., Ph.D. thesis, University of California (1984).
\item[] 
Dickinson, M., private communication (1998).
\item[]
Fort, B., and Mellier, Y., {\it Astron. Astrophys. Rev.}, {\bf 5}, 
239-292 (1994).
\item[] 
Hardcastle, M. J., Lawrence, C. R., and Worrall, D. M., to appear in the 
{\it ApJ} (1998).
\item[]
Stark, A. A., Gammie, C. F., Wilson, R. W., Bally, J., Linke, R.A., 
Heiles, C., \& Hurwitz, M.,
{\it ApJS}, {\bf 79}, 77 (1992).
\item[] 
Turner, T. J., George, I. M., Nandra, K. and Mushotzky, R. F., {\it 
ApJS}, {\bf 113} (1997).
\item[]
Wu, Z-P. \& Fang, L-Z, {\it ApJ}, {\bf 483}, 62 (1997).

\end{itemize}

\end{document}